\documentclass[a4paper,12pt]{article}

\usepackage{setspace}

\usepackage{booktabs}

\usepackage{graphicx}
\usepackage{cite}
\usepackage{amsmath}
\usepackage{amsfonts}
\usepackage{amssymb}
\usepackage{rotating}

\usepackage{caption}
\usepackage{float}

\usepackage{slashed}

\usepackage{hyperref} %links para las referencias

\usepackage[T1]{fontenc}

 \usepackage{lmodern}
 
\usepackage{datetime}
\newdateformat{mydate}{\monthname[\THEMONTH] \THEYEAR} 

\hypersetup{colorlinks=true, linkcolor=black, urlcolor=blue, citecolor=black}

\begin{document}

\begin{titlepage}

\vspace*{-15mm}
\begin{flushright}
TTP18-014 \\
\end{flushright}
\vspace*{0.7cm}

\begin{center} {
\bfseries\LARGE
The mass ratios parametrization
}\\[8mm]
U.~J.~Saldana-Salazar$^{\dagger}$
\quad
and\quad
K.~M.~Tame-Narvaez$^{\star}$
\\[1mm]
\end{center}
\vspace*{0.50cm}
\centerline{$^\dagger$ \itshape
Institut f\"ur Theoretische Teilchenphysik,
Karlsruher Institut f\"ur Technologie,}
\centerline{\itshape Engesserstra{\ss}e 7, D-76131 Karlsruhe, Germany.}
\centerline{$^\star$ \itshape
Institut f\"ur Theoretische Physik, Universit\"at Heidelberg,}
\centerline{\itshape
Philosophenweg 16, D-69120 Heidelberg, Germany.}
\vspace*{1.20cm}

\begin{abstract}	
	The observed hierarchy in the fermion masses,
	which imply a set of small mass ratios, is not naturally small regarding 
	't Hooft's criteria. In this work, in a model independent approach, 
	we introduce a set of conditions
	by which fermion mass ratios become natural. Interestingly, these conditions
	demand that fermion mixing should be described by the four independent
	mass ratios of each fermion sector. 
	Application of this set of conditions to the standard theory
	enables one to understand the mains aspects in quark and lepton mixing. 
	This feature can be taken as a strong evidence for the existence
	in Nature of a flavour symmetry.
	Also, for this analysis to work in the 
	lepton sector, neutrino masses should have normal ordering
	with the lightest neutrino mass satisfying the lower bound, 
	$m_{\nu 1} \geqslant (5.0\pm 0.1) \text{ meV}$,	
	making the approach testable.
\end{abstract}

\end{titlepage}

%\singlespacing
%\tableofcontents
%\singlespacing

%\setcounter{footnote}{0}
%\newpage

\section{Introduction}
	The observed values in the fermion masses together with the ones in quark
	and lepton mixing give origin to very curious patterns,
	\begin{align}
	\begin{split}
		m_t & \sim  \; \Lambda_\text{EW} \;, \\
		m_t & \gg  \; m_f \;, \\
		m_t & \gg \; m_b > m_\tau > m_c \; , \\ 
		m_c & \gg \; m_\mu > m_s \gg m_d \; , \\ 
		m_d & > \; m_u > m_e \;, \\
		m_e & \gg \sum_{\nu} m_{\nu}\; , \\
		m_{3}^2 & \gg \; m_{2}^{2} \gg m_{1}^{2}  \;,
	\end{split}
	\end{align}
	where the last relation holds for each charged fermion species and possibly also for neutrinos, 
	as the quasi-degenerate scenario has been already disfavoured by the cosmological limit on the total sum
	of the three neutrino masses $\sum_\nu m_\nu < 0.23 \text{ eV}$~\cite{Ade:2015xua},
	whereas for the mixing data,
	\begin{align}
	\begin{split}
		{\bf V}_\text{CKM} & = {\bf I} + {\bf \Delta}_q \quad \left( |{\bf \Delta}_{q,ij}| \ll 1 \right) \; , \\ 
		{\bf U}_\text{PMNS} & =  {\bf \Delta_\ell} \quad \left( |{\bf \Delta}_{\ell,ij}| \sim {\cal O}(1) \right)  \; .
	\end{split}
	\end{align}
	This set of yet not understood patterns are known as the problem of masses and mixing, respectively, and are part
	of the \textit{flavour puzzle}, for recent discussions see~\cite{Feruglio:2015jfa,King:2017guk}. In this work,
	we only focus on the problem of mixing. 
	Our discussion here lies in the understanding of the observed values in
	quark and lepton mixing	by virtue of the corresponding fermion masses.

	In the Standard Model (SM), there is no relation between the fermion masses and mixing even though 
	both sets arise from the same source.
	To see this, recall that the Yukawa interactions,
	\begin{equation}
		-{\cal L}_Y \supset {\mathbf{Y}}^{ij}_f \bar{F}_{L,i} \Phi f_{R,j}  + h.c. \; ,
	\end{equation}	 
	with ${\mathbf{Y}}^{ij}_f$ a complex number and entry of a three by three matrix, are parametrised by a large set
	of parameters. After considering the 
	massive nature of neutrinos, altogether 
	they represent 72 (66) low energy parameters in the case of Dirac (Majorana) neutrinos, respectively. 
	However, only twenty (twenty-two) of them can be called physical. In the mass basis, they are represented
	by twelve fermion masses and eight (ten) mixing parameters~\cite{Santamaria:1993ah}. In order to introduce
	our notation, we briefly go through the following well-known steps.
	
	In the initial weak basis, assignment of a non-zero vacuum expectation value to the neutral component of the scalar field,
	\begin{equation}
		\Phi (x) = 
		\begin{pmatrix}
			G^+(x)		\\
			\frac{v+h(x)+iG^0(x)}{\sqrt{2}}
		\end{pmatrix}			\; ,
	\end{equation}		
	spontaneously breaks the electroweak symmetry and brings about the massive nature of fermions,
	\begin{equation}
		\mathbf{M}_f = \frac{v}{\sqrt{2}} \mathbf{Y}_f \; ,
	\end{equation}
	where $v \simeq 246 \text{ GeV}$. Thereafter, diagonalization of the mass matrices,
	\begin{equation} \label{SVD}
		\mathbf{\Sigma}_f = \mathbf{L}_f \mathbf{M}_f \mathbf{R}_f^\dagger \; ,
	\end{equation}
	occurs via a biunitary transformation each acting independently in the left or right-handed corresponding field,	
	\begin{equation}
		F_L\rightarrow \mathbf{L}_f F_L \qquad \text{and} \qquad f_R\rightarrow \mathbf{R}_f f_R \; .
	\end{equation}
	In this new basis, the mass basis, the quark and leptonic charged currents have changed to,
	\begin{equation}
		\mathcal{J}^{\mu,-}_\text{cc-q} = -\frac{g_w}{\sqrt{2}} \bar{u}_L \gamma^\mu \mathbf{V} d_L \; , \qquad
		\mathcal{J}^{\mu,+}_\text{cc-$\ell$} = -\frac{g_w}{\sqrt{2}} \bar{e}_L \gamma^\mu \mathbf{U} \nu_L \; ,
	\end{equation}
	where $\mathbf{V} = \mathbf{L}_u \mathbf{L}_d^\dagger$ and $\mathbf{U} = \mathbf{L}_e \mathbf{L}_\nu^\dagger$. 
	These matrices parametrize how likely are the transitions between 
	any two given unequal flavours via the interactions with the $W^\pm$ bosons. 
	The similarities between these two matrices depends on the massive nature of neutrinos (Dirac or Majorana).
	The quark mixing matrix, $\mathbf{V}$,
	has no unique parametrization,  it is unitary, and it requires of four independent parameters.
	The same could be said about $\mathbf{U}$ when massive neutrinos are introduced as Dirac fermions.
	However, in the case of massive Majorana neutrinos, the mixing matrix is no longer unitary and two more phases
	are needed.\footnote{To understand this, recall that the Majorana condition $\nu^c = \nu$ forbids phase field redefinitions.
	For Dirac neutrinos, the amount of independent complex phases in the mixing matrix is given by $(n-1)(n-2)/2$ where $n$ is
	the number of fermion families.
	However, in the Majorana case, due to the reality condition, the amount of independent complex phases is $n(n-1)/2$.
	Therefore, for $n=3$, one has 1 and 3 non-removable complex phases for Dirac and Majorana neutrinos, respectively.} Conventionally, in the Majorana case, the mixing matrix, $\mathbf{U}$, may be written in the form,
	\begin{align}
		\mathbf{U} = \tilde{\mathbf{U}} \mathbf{K} \;,
	\end{align}
	where $\tilde{\mathbf{U}}$ has all mixing parameters as in the Dirac case with only one non-removable complex phase,
	called the Dirac phase, while $\mathbf{K}$ is a diagonal matrix with the two additional complex phases, 
	\begin{align} \label{eq:MajPhases}
		\mathbf{K} = \text{diag}\left( e^{i \alpha} , e^{i \beta} , 1 \right) \;,
	\end{align}
	called Majorana phases.
	For last, an invariant measure of Charge-Parity ($CP$)
	violation, independent of the parametrization, is the Jarlskog invariant~\cite{Jarlskog:1985ht},
	\begin{align}
		{J}_X = \frac{\text{Im} \left( \text{det} \left[\mathbf{M}_a \mathbf{M}_a^\dagger, \mathbf{M}_b \mathbf{M}_b^\dagger  \right] \right)}{-2\Pi_{i>j }(m_{a,i}^2-m_{a,j}^2) \Pi_{k>l }(m_{b,k}^2-m_{b,l}^2)} \; ,
	\end{align}
	where $X = q, \ell$, $a=u,\nu$, and $b=d,e$. 
	
	There has been many parametrization proposals~\cite{Cabibbo:1963yz,Kobayashi:1973fv,Maiani:1977yd,Schechter:1980gr, 	Wolfenstein:1983yz,Chau:1984fp,Buras:1994ec,Fritzsch:1997st,Charles:2004jd,Rodejohann:2011vc,Hollik:2014jda}. 
	Among them there is one particular parametrization which has served, in the quark sector,
	to provide a better connection to the parameters in flavour physics~\cite{Wolfenstein:1983yz,Buras:1994ec,Charles:2004jd},
	\begin{equation} \label{eq:wolfenstein}
		\mathbf{V}_\text{CKM} = \begin{pmatrix}
			1-\frac{\lambda^2}{2} & \lambda & A\lambda^3(\rho - i\eta) \\
			-\lambda & 1-\frac{\lambda^2}{2} & A\lambda^2 \\
			A\lambda^3(1-\rho - i\eta) & -A\lambda^2 & 1
		\end{pmatrix} + O(\lambda^4) \;.
	\end{equation}		
	The Wolfenstein parametrization, shown in Eq.~\eqref{eq:wolfenstein}, exploits the
	hierarchical structure of the mixing matrix elements, and takes one of them,
	$\lambda \approx 0.22$, as a mixing and expansion parameter along 
	with 
	other three real parameters $A$, $\rho$, and $\eta$ of order ${\cal O}(1)$. 
	This parametrization has been already improved in Ref.~\cite{Buras:1994ec} in order to guarantee unitarity
	of the quark mixing matrix to all orders in $\lambda$.
	
	The standard parametrization, for both quarks and leptons, as suggested by the Particle Data Group 
	(PDG),
	follows Chau and Keung's proposal~\cite{Chau:1984fp},
	\begin{eqnarray}
		\mathbf{W} = \begin{pmatrix}
			1 & 0 & 0 \\ 0 & c_{23} & s_{23} \\ 0 & -s_{23} & c_{23}
		\end{pmatrix}
		\begin{pmatrix}
			c_{13} & 0 & s_{13} e^{-i\delta} \\ 0 & 1 & 0 \\ -s_{13} e^{i\delta} & 0 & c_{13}
		\end{pmatrix}
		\begin{pmatrix}
			c_{12} & s_{12} & 0 \\ -s_{12} & c_{12} & 0 \\ 0 & 0 & 1
		\end{pmatrix},
	\end{eqnarray}
	where we have denoted $c_{ij} = \cos\theta_{ij}$ and $s_{ij} = \sin\theta_{ij}$ and ${\bf W}$ generically represents
	here either quark or lepton mixing for Dirac neutrinos. In the Majorana scenario, mixing with this parametrization
	would be read like $\mathbf{U} = \mathbf{W} \mathbf{K}$ with $\mathbf{K}$ given in Eq.~\eqref{eq:MajPhases}. It 
	has been shown that in the lepton sector a symmetrical parametrization first 
	introduced by Schechter and Valle~\cite{Schechter:1980gr} and later revisited~\cite{Rodejohann:2011vc} 
	gives a similar description but with an additional feature when considering Majorana neutrinos, which is, 
	that the effective mass parameter characterizing the amplitude for neutrinoless double beta decay only 
	depends, as it should, in the two Majorana phases, $\alpha$ and $\beta$, whereas the PDG 
	parametrization also includes the Dirac phase. 
	Furthermore, seesaw extensions of the SM expect deviations from unitarity
	due to the admixture of heavy right-handed neutrinos. For that purpose,
	 an adequate description for non-unitary neutrino mixing has also been
	 proposed~\cite{Escrihuela:2015wra}.

	Mixing parametrizations including mass ratios as mixing parameters have been implicitly~\cite{Hall:1993ni,Fritzsch:1997st,Xing:1996hi,Rasin:1997pn,Chkareuli:1998sa,
    Rasin:1998je,Fritzsch:1999ee,Chkareuli:2001dq,Gerard:2008nc} and explicitly~\cite{Fritzsch:1977vd,Hollik:2014jda} suggested. 
    Nevertheless, only one has really gathered all the four 
	independent fermion mass ratios as mixing parameters in full
	agreement with the observed mixing phenomena~\cite{Hollik:2014jda}. 
	We take here the idea of understanding mixing through masses and investigate it further through a novel approach.
	Of course, although there is no unique way of parametrising, the possibility of
	creating a particular parametrization through fermion mass ratios seems very desirable as this would
	give us a first step into solving one of the biggest old mysteries in Particle Physics.
	Furthermore, the clear advantage of taking mass ratios as mixing parameters compared
	to any other parametrization is that fermion masses are the invariants (singular values) of the
	mass matrices, which in a way, make them more relevant as fundamental parameters for describing mixing
	besides the trivial fact that the number of arbitrary parameters in the weak sector could be significantly 
	reduced from 20 (22) to 12 (14) if the massive nature of neutrinos is also considered as Dirac (or Majorana).

	Moreover, despite the fact that relating mixing to fermion masses has been an active research subject 
	since the pioneer work of Gatto, Sartori, and Tonin~\cite{Gatto:1968ss}, 
	we want to emphasize that our discussion here is novel in many aspects. First, we are
	assuming the viewpoint that if masses ought to explain mixing this should necessarily require 
	having the four independent mass ratios of each fermion sector as 
	mixing parameters without the need to consider other auxiliary parameters, therefore, 
	giving rise to a full mixing parametrization
	as first suggested in Ref. \cite{Hollik:2014jda}. 
	Second, the analysis here made is model-independent and follows a bottom-up approach.
	Third, through 't Hooft's naturalness criteria \cite{tHooft:1979rat} it is pointed out that
	such a new parametrization could implement the hierarchical nature of fermion masses, $m_1^2 \ll m_2^2 \ll m_3^2$,
	and so all generations except the third one would be \textit{naturally} small in the two different limits
	$m_1, m_2 \rightarrow 0$ and $m_1 \rightarrow 0$. Last but not least, without any explicit
	realization of such relations among mixing angles and mass ratios
	quark and lepton mixing can be understood in their main features (small quark mixing
	and anarchical lepton mixing).	
	
	This work is organized as follows. In Section \ref{sec:general}, 
	we briefly introduce the main general aspects when establishing a relation between mixing angles
	and mass ratios. Next, in Section \ref{sec:naturalness}, 
	we show how the hierarchical nature of fermion masses if seen as natural demands the mixing matrices
	to be written in terms of the fermion masses and satisfying different mass limits. 
	Thereafter, in Section \ref{sec:1stApproach}, we assume the corresponding
	four independent mass ratios of each sector as the mixing parameters without any explicit
	functional realization. By sole virtue of it, we study quark and lepton mixing 
	through the phenomenological input of hierarchical fermion masses.  	
	Finally, in Section \ref{sec:Concl}, we conclude.

	\section{General aspects}
	\label{sec:general}

	The usual procedure to reparametrize comes from the matrix invariants which are the coefficients
	of the characteristic polynomial, $\text{det}[\mathbf{M}_f\mathbf{M}_f^\dagger-\lambda \mathbf{I}] = 0$. In the $n$ family case, the set of $n$ invariants provides $n$
	equations which can be used, although not always easily, to write the matrix parameters in terms
	of the singular values (masses). For simplicity, the $n=3$ case would be given by,
	\begin{equation}
		\lambda^3 -  \text{tr}[\mathbf{H}_f ] \lambda^2 + \frac{1}{2} \left(\text{tr}[\mathbf{H}_f]^2 - \text{tr}[\mathbf{H}_f \mathbf{H}_f] \right) \lambda - \text{det}[\mathbf{H}_f] = 0 \; ,
	\end{equation}
	where $\mathbf{H}_f = \mathbf{M}_f \mathbf{M}_f^\dagger$ is the hermitian product
	and the roots of the equation are the eigenvalues (squared masses) of $\mathbf{H}_f$. 
	
	The matrix invariants in terms of the	masses are written as,
	\begin{align}
		 \text{tr}[\mathbf{H} ] & = m^2_1 + m^2_2 + m_3^2 \; , \\
		 \text{det}[\mathbf{H}] & = m^2_1  m^2_2  m_3^2 \; , \\
		 \frac{1}{2} \left(\text{tr}[\mathbf{H}]^2 - \text{tr}[\mathbf{H} \mathbf{H}] \right) & =
		 m^2_1  m^2_2  +   m^2_2  m_3^2 + m^2_1   m_3^2 \; ,
	\end{align}
	where in these last expressions we have suppressed the subscript $f$ as this applies to all fermions. 

	Let us consider the Weinberg ansatz~\cite{Weinberg:1977hb} to provide a simple example, 
	\begin{equation}
		\mathbf{m} = 
		\begin{pmatrix}
			0 & |a| \\
			|a| & |b|			
		\end{pmatrix} \qquad \rightarrow \qquad 
		\begin{matrix}
		 \text{tr}[\mathbf{m} ] = |b|= m_1 + m_2 \\
		 \text{det}[\mathbf{m}] = -|a|^2   = m_1 m_2
		\end{matrix} \qquad \rightarrow \qquad
		\begin{pmatrix}
			0 & \sqrt{m_1 m_2} \\
			\sqrt{m_1 m_2} & m_2 - m_1			
		\end{pmatrix} \; ,
	\end{equation}
	where we have considered without any loss of generality $m_1 \rightarrow -m_1$ \footnote{This change of sign can be easily achieved by a 
	global chiral transformation.}.
	 In return, we immediately
	obtain a relation between the angle of rotation and a mass ratio, $\tan \theta = \sqrt{m_1 /m_2}$. In fact, this ansatz was made to reproduce the well known Gatto--Sartori--Tonin relation for the Cabibbo angle~\cite{Gatto:1968ss}, $\theta_C \approx \sqrt{m_d/m_s}$. We must add another remark to this example, even though we have two different masses, we can
	always consider the largest mass as setting the scale of the matrix whereas the ratio with the 
	lighter one the relevant parameter, as it provides all the internal
	structure of the mass matrix,
	\begin{equation} \label{eq:weinberg}
		\mathbf{m} = m_2 \, \begin{pmatrix}
			0 & \gamma\\
			\gamma & 1 - \gamma^2			
		\end{pmatrix} =
		m_2 \, \begin{pmatrix}
			0 & 0\\
			0 & 1
		\end{pmatrix} +
		m_2 \, \begin{pmatrix}
			0 & \gamma \\
			\gamma & - \gamma^2			
		\end{pmatrix}		 \; ,
	\end{equation}
	where we defined it as $\gamma \equiv \sqrt{\frac{m_1}{m_2}}$. In fact, due to the hierarchy in the masses
	this parameter is expected to be small.
	Explaining its smallness belongs to the problem of masses. For example,
	it could be originated in a model where the lighter masses arise from radiative
	corrections~\cite{Ibarra:2014pfa,Ibarra:2014fla,CarcamoHernandez:2016pdu}, from 
	extra dimensions~\cite{ArkaniHamed:1999dc} or the Froggatt--Nielsen mechanism~\cite{Froggatt:1978nt}.
	On the other hand, already at this stage, two indispensable aspects can be realized 
	from Eq.~\eqref{eq:weinberg}: null matrix elements,  ${\bf M}_{ij} = 0$, and relations among matrix
	elements, ${\bf M}_{ij} = {\bf M}_{k l}$. Both requirements can be met by several approaches,
	for example, texture-zeros~\cite{Fritzsch:1977vd, Weinberg:1977hb, Branco:1988iq,
  Ramond:1993kv, Branco:1994jx, Branco:1999nb, Branco:2010tx,
  Emmanuel-Costa:2016gdp}, flavour symmetries~\cite{Ishimori:2010au}, reinterpretation
	of the fermion mass matrix elements~\cite{Hollik:2017get}, among others.

	Regarding the feasibility of a reparametrisation, let us discuss when it is possible.
	A complex $n \times n$ matrix has $n^2$ phases and $n^2$ magnitudes.
	By virtue of the $n$ invariants, and without further constrains, the task of reparametrising in terms of the masses 
	is impossible as the system is underdetermined.  	
	Of course, a further reduction of the arbitrariness is still possible if we recall that the kinetic terms
	\textit{per} fermion sector posses 
	a $[U(n)]^3$ due to the universality of the gauge couplings. This accidental symmetry 
	group describes the nature of the transformations leaving invariant the
	weak interaction basis\footnote{By weak interaction basis, we mean those bases where
	the weak interactions are diagonal in flavour space.}. 
	We have at our disposal: $\tfrac{3n(n-1)}{2}$ 
	and $\tfrac{3n(n+1)-2}{2}$ arbitrary magnitudes and complex phases, respectively,
	to choose whatever basis we require. As these transformations are involved in
	the two kinds of fermions of a given sector, we continue our counting by summing up all
	the parameters of the corresponding two mass matrices: $2n^2$ magnitudes and $2n^2$
	complex phases. A careful choice of basis, with both mass matrices still not fully diagonal,
	would have $\tfrac{n(n+3)}{2}$ and $\tfrac{(n-1)(n-2)}{2}$ arbitrary parameters in
	magnitudes and complex phases, respectively. Reparametrization with the $2n$ invariants
	would still leave $(n-1)^2$ arbitrary parameters. In particular, for $n=3$, this means 
	that there are \textit{special}
	bases where the mass matrices can be reexpressed in terms of its singular values plus four
	unknown \textit{physical} parameters.  In fact, these parameters should be equivalent to 
	the mixing parameters appearing in the mass basis. 
	An example of such a basis would be the following couple of matrices,
	\begin{equation}
		\mathbf{M}_a = 
		\begin{pmatrix}
			m_{1}^a & 0 & 0\\
			0 & m_{2}^a & 0\\
			0 & 0 & m_{3}^a		
		\end{pmatrix}, \; \qquad
		\mathbf{M}_b = 
		\begin{pmatrix}
			m_{11}^b & 0 & 0\\
			m_{21}^b & m_{22}^b e^{-i \delta} & 0\\
			m_{31}^b & m_{32}^b & m_{33}^b		
		\end{pmatrix} \; ,
	\end{equation}
	where we have employed Ref.~\cite{Hollik:2017get} to find such a basis. In this case, one finds that, 
	\begin{align}
		\begin{split}
		m_{33}^b & = m_3^b \;, \\
		m_{11}^b m_{22}^b & = m_1^b m_2^b \; , \\
		(m_{11}^b)^2 + (m_{22}^b)^2 + (m_{21}^b)^2 + (m_{31}^b)^2 + (m_{32}^b)^2 & = 
		(m_1^b)^2 + (m_2^b)^2  \; .
		\end{split}
	\end{align}
	Further reduction, should only be possible within an ultraviolet completion of the SM 
	in which the problem of mixing gets a solution. 
		
	For last, to fully reparametrise a mass matrix in terms of its singular values
	the number of independent mass ratios should be larger or equal than the number of mixing parameters.
	In fact, this only occurs for two and three fermion generations \cite{Hollik:2014jda}. 
	To see this, assume again $n$ fermion
	generations. The number of independent mass ratios, $2(n-1)$, grows much slower than 
	the number of mixing parameters, $(n-1)^2$. As a consequence, 
	being able to completely reparametrize depends
	on the inequality $2(n-1) \geq (n-1)^2$ and thus $1< n \leq 3$. Although this might just be
	an accident, it is an interesting coincidence that one may express angles by mass ratios
	not for an arbitrary number of generations. One could try to read it as another hint for an underlying
	more fundamental theory of flavour.

	\section{Naturalness and hierarchical fermion masses}
	\label{sec:naturalness}

A small number is natural only if an exact symmetry emerges when it is set to zero. This is 't Hooft's criteria for naturalness~\cite{tHooft:1979rat}. Although nowadays this criteria and its usefulness as a guiding principle is being
questioned~\cite{Dine:2015xga,Giudice:2017pzm}, here we will still consider it as valid and find out an application to the problem of mixing. 

It is not new that the hierarchy in the quark sector, both in the masses and mixing, might be a hint
of a possible connection among them. In this regard, the smallness in fermion masses could be
understood as a consequence of symmetries being approximately 
conserved~\cite{Antaramian:1992ya,Hall:1993ca,Leurer:1993gy,Gerard:2008nc}.
However, this insight gets weakened when considering the lepton sector with its anarchical mixing. In the following, we show how the hierarchical nature among all fermion masses, 
\begin{align} \label{eq:hierarchy}
	m_1^2 \ll m_2^2 \ll m_3^2 \; ,
\end{align}
described by just two ratios $\tfrac{m_2^2}{m_3^2}$ and $\tfrac{m_1^2}{m_2^2}$, 
if related to approximately conserved flavour symmetries,
strictly implies relations between fermion masses and mixing to the fullest extent.

In the weak interaction basis, 
the SM lagrangian acquires an exact global symmetry when all Yukawa couplings are set to zero, 
\begin{align}
	\mathcal{G}_F = U_L^Q(3) \times U_R^u(3) \times U^d_R(3) \times U_L^E(3) \times U_R^e(3) \; .
\end{align}
In this respect, the smallness of the set of all Yukawa couplings is natural. In the mass basis,
this means that all fermion masses are much smaller than the flavour scale,
${m_f} \ll {\Lambda_F}$, wherein the theory of flavour is expected to be realized, ${\Lambda_F} \gtrsim {\cal O}(1 \text{ TeV})$.

On the other hand, we may also ask how natural is the hierarchy in the fermion masses, Eq. \eqref{eq:hierarchy},
such that when either $m_1 , m_2 \rightarrow 0$ or $m_1 \rightarrow 0$ there emerges
an exact symmetry. In the SM, the fermion mass hierarchy is not natural. Let us see this. The Yukawa matrices, in the weak basis, break the flavour group to,
\begin{align}
	\mathcal{G}_F \xrightarrow[{\bf Y}_f]{} U_{B}(1) \times U_e(1) \times U_\mu(1) \times U_\tau(1) \; ,
\end{align}
where $B$ denotes baryon number.
From Eq. \eqref{SVD}, we have,
\begin{align}
	\widetilde{\bf M} = \frac{{\bf M}}{m_3} = {\bf L}^\dagger \begin{pmatrix}
		\frac{m_1}{m_3} & 0 & 0 \\
		0 & \frac{m_2}{m_3} & 0 \\
		0 & 0 & 1
	\end{pmatrix}
	{\bf R} \; ,
\end{align} 
where, for convenience, we have instead used the mass matrix
normalized by its largest singular value and omitted the fermion type index. In any
of the two limits there is no symmetry emerging. Take for example, the simplest case
with $m_1 , m_2 \rightarrow 0$, we get the rank one matrix,
\begin{align}
	\widetilde{\bf M} \simeq \begin{pmatrix}
		L^*_{31} R_{31} & L^*_{31} R_{32} & L^*_{31} R_{33} \\
		L^*_{32} R_{31} & L^*_{32} R_{32} & L^*_{32} R_{33} \\
		L^*_{33} R_{31} & L^*_{33} R_{32} & L^*_{33} R_{33}
	\end{pmatrix} \; .
\end{align} 
The unitary matrices diagonalizing both rank one matrices (either in the quark or lepton sector)
will be in general independent from each other and therefore their product cannot be expected to
be the unit matrix, for more details see Appendix~\ref{app:Basis}. 
Then, we must expect fermionic mixing even in the lowest rank scenario.
And a similar situation for $m_1 \rightarrow 0$. Of course this is a basis
dependent observation as one may have a certain basis wherein both rank one matrices could be diagonalized by
the same unitary transformation and thus no mixing may appear (e.g. the democratic scenario).
Regarding this observation, without any loss of generality, we now choose to work in the mass basis and also introduce the massive nature of neutrinos by assuming, for simplicity, Dirac neutrinos (the Majorana case would at most
have an orthogonal global symmetry group instead of a unitary one in ${\cal G}_F$). 

In the mass basis, there are no right-handed flavour changing processes. 
Therefore, making the first two generations
massless implies, $\{ m_1, m_2\} \rightarrow 0$,
\begin{align} \label{eq:U2}
	 U_R^a(2) \times U^b_R(2) \; ,
\end{align}
where  $a=u,\nu$ and $b=d,e$. 
This is in contradiction to the conclusion obtained in the weak basis, as no symmetry was found there. 
To avoid the contradictions we need  to obtain a set of basis-invariant conditions in which hierarchical fermion masses become always \textit{natural}.

Let us continue in the mass basis. In a way, we could say the lightness of the first two generations compared to the third one is natural or at least
partially natural, as there is no symmetry
corresponding to the left-handed fields. In fact, the emergent symmetry would be maximal if they also had it. 
Similarly, if we make the first generations massless,
\begin{align} \label{eq:U1}
	 U_R^a(1) \times U^b_R(1) \; ,
\end{align}
no symmetry appears for the left-handed fields. 
There is clearly an issue in this picture where only the right-handed parts get new symmetry
factors~\cite{Antaramian:1992ya}. The underlying theory must have approximate
flavour symmetries acting on both the left- and right-handed fields. Otherwise,
we would have no reason to assign $t_L$ and $b_L$ into the same
$SU(2)_L$ doublet~\cite{Antaramian:1992ya}. 

Now, there is a possibility to solve this issue if we require the following. 
If, in the mass basis, for a given fermion sector, $F=Q,E$,
\begin{equation} \label{eq:LagConditions}
	{\cal L} \supset \overline{F}_{L,a} {\bf W} \gamma^\mu F_{L,b} W_\mu - \sum_a m_a \overline{F}_{L,a} f_{R,a} - \sum_b m_b \overline{F}_{L,b} f_{R,b} + \text{ H.c} \; ,
\end{equation}
we demand the mixing matrix, from here onwards \textit{generically} denoted by $\mathbf{W}$, to satisfy the functional dependence,
\begin{equation} \label{eq:cond1}
	{\bf W} = {\bf W} \left( \frac{m_1^a}{m_2^a}, \frac{m_2^a}{m_3^a}, \frac{m_1^b}{m_2^b}, \frac{m_2^b}{m_3^b} \right) \; ,
\end{equation}
and fulfill the two limits,
\begin{itemize}
	\item $\{m_1, m_2\} \rightarrow 0$:
	\begin{equation}\label{eq:cond2}
 {\bf W} \left( 0, 0, 0, 0 \right) = {\bf I} \; ,
	\end{equation}
	implying the global flavour symmetry group or any subgroup contained within it,
	\begin{equation}
	U_L^F(2) \times U_R^a(2) \times U^b_R(2) \; ,
	\end{equation}
	and where the limit should be taken having in mind that $m_1 \ll m_2$.

	\item $\{m_1\} \rightarrow 0$:
	\begin{equation}\label{eq:cond3}
 {\bf W} \left( 0, \frac{m_2^a}{m_3^a}, 0, \frac{m_2^b}{m_3^b} \right) = {\bf W}_{23}\left( \frac{m_2^a}{m_3^a}, \frac{m_2^b}{m_3^b} \right) \; ,
	\end{equation}
	implying the global flavour symmetry group or any subgroup contained within it,
	\begin{equation}
		U_L^F(1) \times U_R^a(1) \times U^b_R(1) \; ,
	\end{equation}
	and here ${\bf W}_{23}$ represents a unitary transformation acting only in the 2-3 family subspace. 
\end{itemize}
where in Eq.~\eqref{eq:LagConditions} $f_R$ or $F_L$ denote either a weak singlet or a weak doublet, respectively,
the two doublet components have appeared explicitly, $F_L = (F_{L,a},F_{L,b})^T$,
and represent in family space, together with the right-handed parts, three dimensional vectors, for example, $F_{L,a} = \left( F_{L,a1}, F_{L,a2}, F_{L,a3} \right)^T$,  
${\bf W} = {\bf V}, {\bf U}$, $F=Q,E$, $a=u,\nu$, and $b=d,e$, respectively.
Then, under these conditions, symmetries do emerge for both handedness in the two independent
cases and fermion masses could be regarded as fully natural. As a corollary, given the previous limits, we infer the
proposed parametrization should intrinsically satisfy also the decoupling limit, $m_3 \rightarrow \infty$,
\begin{equation} \label{eq:cond4}
{\bf W} \left( \frac{m_1^a}{m_2^a}, 0, \frac{m_1^b}{m_2^b},0 \right) = {\bf W}_{12}\left( \frac{m_1^a}{m_2^a},\frac{m_1^b}{m_2^b}\right) \; .
\end{equation}

Hence, if the first two generations are naturally small compared to the third generation
and the first generation naturally small compared to the second generation this would then
mean having a mass ratios mixing parametrization satisfying the necessary limits in such a 
way that a global flavour symmetry could be separately recovered in each case. 

It could naively seem that these conditions are arbitrary as they were not derived from special texture
zeros in the mass matrices or by assigning fermion fields to particular irreducible representations of the
corresponding global flavour symmetry groups. Nevertheless,
these conditions point to a concrete class of models where naturalness should be realized as shown in Table \ref{table-limits},
 in agreement
to the conditions expressed by Eqs. \eqref{eq:cond1}, \eqref{eq:cond2}, and \eqref{eq:cond3}.
These conditions point to the sequential breaking of the maximum flavour symmetry,
\begin{equation}
	[U(3)]^6 \quad \xrightarrow[\{m_3\}]{} \quad [U(2)]^6 \quad \xrightarrow[\{m_2\}]{} \quad [U(1)]^6 \quad \xrightarrow[\{m_1\}]{}\quad U(1)_{B}\times U(1)_{L} \; ,
\end{equation}
 where
intermediate trivial $U(1)$ factors are left for readability. 
For example, models with minimally broken flavour symmetry~\cite{Barbieri:1996ww,Barbieri:1997tu,Blankenburg:2012nx,Buras:2012sd,Barbieri:2012uh} where the approximate $U(2)$ flavour symmetry is mainly employed. Its application can also be found in the case of supersymmetric~\cite{Barbieri:1995uv,Crivellin:2008mq,Crivellin:2011sj} and grand unified~\cite{Linster:2018avp} theories. The meaning of $U(2)$ flavour symmetries which may be used in a weaker symmetry assignment~\cite{Aranda:1999kc,Linster:2018avp}, is the
arrangement of the first two generations into one doublet whereas the third one transforming as a singlet. 

\begin{table}
\centering
\begin{tabular}{cc}
\hline \hline
Fermion mass limits & Emergent symmetry\\
\hline \hline
 $\{m_1,m_2,m_3\}\rightarrow 0$ & $ {\cal S}^{(3)}_F \subseteq [U(3)]^6$  \\
\hline
$\{m_1,m_2\}\rightarrow 0$ & ${\cal S}^{(2)}_F \subseteq[U(2)]^6$  \\
\hline
$\{m_1\}\rightarrow 0$ & ${\cal S}^{(1)}_F \subseteq [U(1)]^6$  \\
\hline
\hline
\end{tabular}
 \caption{The left column shows the three different cases when the set of all masses, $\{ m_j \}$, from the $j$-generation within a given fermion sector are set to zero. Whereas the right column shows the global flavour symmetry the full lagrangian
will acquire after taking the corresponding limit. Of course, we are considering also the possiblity of having discrete or continous subgroups as denoted by ${\cal S}^{(k)}_F$.}
 \label{table-limits}
\end{table}

For last, the approach is, by construction, only consistent with the normal ordering case which happens to be
in agreement to the most recent global analysis which favours normal ordering over the inverted one at more
than $3\sigma$~\cite{deSalas:2017kay},
therefore, the inverted ordering for neutrino masses will not be considered in the remaining part of the work.

\section{Implications of mass ratios as mixing parameters}
\label{sec:1stApproach}
	The new given dependence on the mass ratios has an immediate consequence: both fermions
	within a sector must contribute to fermion mixing. That is, the departing
	weak interaction basis
	should have all matrices as non-diagonal; and hence, the unitary transformations
	acting in the left-handed fields and diagonalizing the mass matrices,
	\begin{align} 
	\begin{split}
		\mathbf{L}_u \mathbf{M}_u \mathbf{M}_u^\dagger \mathbf{L}_u^\dagger = \mathbf{\Sigma}^2_u \; ,
		\qquad
		\mathbf{L}_d \mathbf{M}_d \mathbf{M}_d^\dagger \mathbf{L}_d^\dagger = \mathbf{\Sigma}^2_d \; , \\
		 \mathbf{L}_e \mathbf{M}_e \mathbf{M}_e^\dagger \mathbf{L}_e^\dagger = \mathbf{\Sigma}^2_e \; ,
		\qquad
		\mathbf{L}_\nu \mathbf{M}_\nu \mathbf{M}_\nu^\dagger \mathbf{L}_\nu^\dagger = \mathbf{\Sigma}^2_\nu \; , 	
	\end{split}
 	\end{align}
 	shall give the desired dependence on the four mass ratios,
 	\begin{align} \label{eq:basis1}
 		\mathbf{V}\left( \frac{m_u}{m_c}, \frac{m_c}{m_t},
		 \frac{m_d}{m_s}, \frac{m_s}{m_b}  \right) & = 
		 \mathbf{L}_u \left( \frac{m_u}{m_c}, \frac{m_c}{m_t} \right) 
		 \mathbf{L}_d^\dagger \left( 
		 \frac{m_d}{m_s}, \frac{m_s}{m_b}  \right) \; ,
 	\\ \label{eq:basis2}
 		\mathbf{U}\left( \frac{m_e}{m_\mu}, \frac{m_\mu}{m_\tau},
		 \frac{m_{\nu 1}}{m_{\nu 2}}, \frac{m_{\nu 2}}{m_{\nu 3}}  \right) & = 
		 \mathbf{L}_e \left( \frac{m_e}{m_\mu}, \frac{m_\mu}{m_\tau}  \right)
		 \mathbf{L}_\nu^\dagger \left( 
		 \frac{m_{\nu 1}}{m_{\nu 2}}, \frac{m_{\nu 2}}{m_{\nu 3}}  \right) \; .
 	\end{align}
	Eqs. \eqref{eq:basis1} and \eqref{eq:basis2} point to a special weak basis the flavour model
	should possess wherein both mass matrices can contribute to mixing. 
	
	Realize how the conditions of Eqs. \eqref{eq:cond1}, \eqref{eq:cond2}, \eqref{eq:cond3}, and \eqref{eq:cond4}
	in any weak basis should imply the following properties for the mass matrices\footnote{In the case of a weak basis with one mass matrix already diagonal, is the non-diagonal one which acquires the mentioned forms.},
	\begin{align}
		\{m_1, m_2\} \rightarrow 0  & &
		\begin{pmatrix}
		0 & 0 & 0 \\
		0 & 0 &  0 \\
		0 & 0 & \blacksquare \\
		\end{pmatrix} \; , \\
		\{m_1\} \rightarrow 0  & &
		\begin{pmatrix}
		0 & 0 & 0 \\
		0 & \blacksquare & \blacksquare \\
		0 & \blacksquare & \blacksquare \\
		\end{pmatrix}	 \; , \\
		\{m_3\} \rightarrow \infty  & &
		 \begin{pmatrix}
		\blacksquare & \blacksquare & 0 \\
		\blacksquare& \blacksquare & 0 \\
		0 & 0 & \blacksquare \\
		\end{pmatrix}		 \; . 
	\end{align}
	Therefore, the hierarchical nature of fermion masses mean the following building process for 
	the fermion mass and mixing matrices\footnote{For an example where such kind of matrix structures can be produced see Ref.~\cite{Saldana-Salazar:2015raa}. },
	\begin{align} \label{eq:MatrixBuilding}
		\begin{split}
       \begin{pmatrix}
		0 & 0 & 0 \\
		0 & 0 &  0 \\
		0 & 0 & 0 \\
		\end{pmatrix}
		 \xrightarrow{{m_3}}
		\begin{pmatrix}
		0 & 0 & 0 \\
		0 & 0 &  0 \\
		0 & 0 & \blacksquare \\
		\end{pmatrix}
		 \xrightarrow{{m_2}}
		  \begin{pmatrix}
		0 & 0 & 0 \\
		0 & \blacksquare & \blacksquare \\
		0 & \blacksquare & \blacksquare \\
		\end{pmatrix}		  	
		\xrightarrow[\mathbf{L}_{23}]{}
		  \begin{pmatrix}
		0 & 0 & 0 \\
		0 & \blacksquare & 0 \\
		0 & 0 & \blacksquare \\
		\end{pmatrix}		
		\xrightarrow{{m_1}} \\
		  \begin{pmatrix}
		\blacksquare & \blacksquare & \blacksquare \\
		\blacksquare& \blacksquare & \blacksquare \\
		\blacksquare & \blacksquare & \blacksquare \\
		\end{pmatrix}
		\xrightarrow[\mathbf{L}'_{23}]{}	
		  \begin{pmatrix}
		\blacksquare & \blacksquare & \blacksquare \\
		\blacksquare& \blacksquare & 0 \\
		\blacksquare & 0 & \blacksquare \\
		\end{pmatrix}	
		\xrightarrow[\mathbf{L}_{13}]{}
		  \begin{pmatrix}
		\blacksquare & \blacksquare & 0 \\
		\blacksquare& \blacksquare & 0 \\
		0 & 0 & \blacksquare \\
		\end{pmatrix}	
		\xrightarrow[\mathbf{L}_{12}]{}
		  \begin{pmatrix}
		\blacksquare & 0 & 0 \\
		0& \blacksquare & 0 \\
		0 & 0 & \blacksquare \\
		\end{pmatrix}
		\end{split}
		\end{align}
	and from it, we can infer, as explicitly shown there, how the unitary transformation acting in the left-handed fields
	should be given by three successive rotations,
	\begin{equation}
		{\bf L}_f = {\bf L}_{12}^f {\bf L}^f_{13} {\bf L}^f_{23} \; ,
	\end{equation}
	where ${\bf L}^f_{ij}$ represents a transformation acting only in the subspace $i-j$ of family space and in Eq.~\eqref{eq:MatrixBuilding} we have considered that turning on the mass of the first family could in general contribute
	to all the matrix elements.
	
	Hence, quark and lepton mixing should be given as,
	\begin{align}
		{\bf V} = {\bf L}_{12}^u {\bf L}^u_{13} {\bf L}^u_{23}{\bf L}^d_{23}{}^\dagger {\bf L}^d_{13}{}^\dagger {\bf L}_{12}^d{}^\dagger \; ,
	\end{align}
	and
	\begin{align}
		{\bf U} = {\bf L}_{12}^e {\bf L}^e_{13} {\bf L}^e_{23}{\bf L}^\nu_{23}{}^\dagger {\bf L}^\nu_{13}{}^\dagger {\bf L}_{12}^\nu{}^\dagger\; .
	\end{align}

	\subsection{The Cabibbo--Kobayashi--Maskawa matrix}
	The general features characterizing quark mixing, as shown in Appendix \ref{app:Mixing},
	can be summarized by two main aspects: small mixing,
	\begin{align}
		\theta^\text{CKM}_{ij} \ll 1 \; ,	
	\end{align}
	and hierarchical mixing,
	\begin{align}
		\theta^\text{CKM}_{12} \gg \theta^\text{CKM}_{23} \gg \theta^\text{CKM}_{13} \; .
	\end{align}		
	Understanding quark mixing would necessarily mean explaining these general features.
	
	In the following, we will apply the mass ratios parametrization (MRP) and 
	exploit the phenomenological observation that all quark 
	masses fulfill the hierarchy\footnote{A somewhat similar approach can be found in Ref. \cite{Antusch:2009hq} wherein
	contributions to the mixing angles coming from the different fermion species were considered small and thereof
	quark and lepton mixing sum rules were obtained. However, here we justify the smallness through
	the smallness of the mass ratios.},
	\begin{align}
		m_3^2 \gg m_2^2 \gg m_1^2 \; .
	\end{align} 		
	
	We denote by $\Theta_{ij}^f$ the angle appearing in the transformation ${\bf L}_{ij}^f$.\footnote{
	Realize that, without appealing to any specific model, we could guess the functional form of this individual
	mixing angle acting in only one fermion type. $\Theta_{ij}^f$ needs to behave as $\Theta_{ij}^f \rightarrow 0$ whenever either $m_i^f \rightarrow 0$ or  $m_j^f \rightarrow \infty$.
	Therefore, the following simple kind of relation,
	\begin{equation}
		\Theta_{ij}^f \sim \left( \frac{m_i^f}{m_j^f} \right)^n  g(\frac{m_k^f}{m_{l}^f})\;,
	\end{equation}
	where $n \in \Re$, $g(\frac{m_k^f}{m_{l}^f})$ is an unknown function that represents the possible 
	contributions from the other two mass ratios, and $k$ and $l$ denote family number, 
	may give the necessary structure for a full analysis. As this analysis is beyond the scope of this work, it is left
	for future work.
	}
	As naturalness indicates, the mixing parametrization in Eq.~\eqref{eq:cond1} should satisfy three limits, see Eqs.~\eqref{eq:cond2}, \eqref{eq:cond3}, and \eqref{eq:cond4}. 
	
	The first aspect of quark mixing can be understood by simply applying Eq.~\eqref{eq:cond2}. Thus, we expect
	quark mixing to have the general form,
	\begin{align}
		{\bf V}_\text{CKM} & \approx 
		\begin{pmatrix}
		1  & 0 & 0 \\
		0  & 1  & 0 \\
		0 & 0 & 1
		\end{pmatrix} + {\cal O}(\theta_{ij}^\text{CKM})\;.
	\end{align}
	
	Now, to understand the second aspect of quark mixing that is, hierarchical mixing, we do the following.
	As ratios coming from the up-quark sector are all negligible compared to 
	the ones in the down sector we do not expect them to significantly contribute
	to quark mixing and therefore are neglected. 
	On the other hand, as $m_d/m_s$ is much larger than $m_s/m_b$ the first dominant contribution in quark mixing 
	should come from the limit
	$m_b \rightarrow \infty$ given in Eq. \eqref{eq:cond4},
	\begin{align}
		\mathbf{V}_\text{CKM} &\approx 		
		\begin{pmatrix}
		1  & -\Theta_{12}^d  & 0 \\
		\Theta_{12}^d  & 1  & 0 \\
		0 & 0 & 1
		\end{pmatrix} \;.
	\end{align}
	
	To include further corrections, we may assume, as suggested from the mass ratios, that $\Theta_{23}^d \sim (\Theta_{12}^d)^2$ and obtain to third order\footnote{Realize that if one 
	takes $\Theta_{23}^d  \approx (\Theta_{12}^d)^2 = \lambda^2$ one can approximately reproduce the well established 
	Wolfenstein parametrization to order $\mathcal{O}(\lambda^3)$. Note that, inside this approach, the $|V_\text{ub}|$ element has a hierarchy of $\mathcal{O}(\lambda^4)$
	instead of the conventional  $\mathcal{O}(\lambda^3)$. This represents no problem as in fact
	if all the Wolfenstein mixing parameters except for $\lambda$ were rescaled one should expect 
	$V_\text{ud} \sim \lambda \approx 0.22$, $V_{cb} \sim \lambda^2 \approx 0.05$, $V_{tb} \sim \lambda^3 \approx 0.01$, and $V_{ub} \sim \lambda^4 \approx 0.002$
	\cite{Hou:1994sa, Hou:1997iv}.}, 
	\begin{align}
	\begin{split}
		\mathbf{V}_\text{CKM} &\approx
		\begin{pmatrix}
		1 & 0 & 0 \\
		0 & 1 -\frac{(\Theta_{23}^d)^2}{2}  & -\Theta_{23}^d  \\
		0 & \Theta_{23}^d  & 1 -\frac{(\Theta_{23}^d)^2}{2} \\
		\end{pmatrix}		
		\begin{pmatrix}
		1-\frac{(\Theta_{12}^d)^2}{2}  & -\Theta_{12}^d + \frac{(\Theta_{12}^d)^3}{3!} & 0 \\
		\Theta_{12}^d -\frac{(\Theta_{12}^d)^3}{3!}   & 1-\frac{(\Theta_{12}^d)^2}{2}  & 0 \\
		0 & 0 & 1
		\end{pmatrix} \; , \\
		 & \simeq
		\begin{pmatrix}
		1-\frac{(\Theta_{12}^d)^2}{2}  & -\Theta_{12}^d + \frac{(\Theta_{12}^d)^3}{3!}  & 0 \\
		\Theta_{12}^d -\frac{(\Theta_{12}^d)^3}{3!}   & 1-\frac{(\Theta_{12}^d)^2}{2}  & -\Theta_{23}^d  \\
		\Theta_{23}^d  \Theta_{12}^d  & \Theta_{23}^d  & 1
		\end{pmatrix} \; ,
		\end{split}
	\end{align}
	where we have Taylor expanded. 
	We may now identify the following mixing sum rules to first order\footnote{The study of $\theta^\text{CKM}_{13}$ is
	beyond the scope of this work, as here we are mainly interested in understanding the main aspects of quark and lepton 
	mixing.},
	\begin{equation}
		\Theta_{12}^d \simeq \theta^\text{CKM}_{12} \qquad \text{ and } \qquad
		\Theta_{23}^d \simeq \theta^\text{CKM}_{23} \; ,
	\end{equation}
	 in agreement to the more general form given in \cite{Antusch:2009hq}.	
	 This is telling us that the dominant contributions producing the observed values
	 in quark mixing come from the down-quark sector and given the hierarchy
	 among its mass ratios, 
	 \begin{equation}
	 	\frac{m_d^2}{m_s^2} \gg \frac{m_s^2}{m_b^2} \gg \frac{m_d^2}{m_b^2} \; ,
	 \end{equation}
	 we should, in general, expect hierarchical mixing obeying the same order,
	 $\theta_{12}^\text{CKM} \gg \theta_{23}^\text{CKM} \gg \theta_{13}^\text{CKM}$.
	 
	Hence, we see the great advantages of employing the MRP as we 
	now roughly understand with very little effort how  the hierarchy in the Cabibbo--Kobayashi--Maskawa 
	(CKM) matrix 
	is indeed a direct consequence of the strong hierarchy in the quark masses. 
	In this sense, the smallness of $m_u/m_t$ and $m_d/m_b$ compared to the other ratios,
	and the fact that $|V_\text{ub}|$ is also observed to be the smallest element in the mixing matrix 
	supports this conclusion.
		
	\subsection{The Pontecorvo--Maki--Nakagawa--Sakata matrix} \label{subsec:app2-PMNS}
	The general features describing lepton mixing, as shown in Appendix \ref{app:Mixing}, 
	can be summarized by the following situation: anarchical mixing,
	\begin{equation}
		|{\bf U}_{\text{PMNS},ij}| \sim {\cal O}(1) 	\; ,
	\end{equation}
	with a very particular hierarchy among the mixing angles,
	\begin{align} \label{eq:LepMixingHierarchy}
		\frac{\pi}{4} \approx \theta_{23}^\text{PMNS} > \theta_{12}^\text{PMNS} > \theta_{13}^\text{PMNS} \sim \theta_{12}^\text{CKM} \; ,
	\end{align}		 
	where we have made explicit how the atmospheric mixing angle is almost maximal whereas the reactor
	one is of the order of the largest element in the CKM mixing matrix.  
	
	Let us apply the MRP to the lepton sector. 	
	For the sake of illustration, in the following we do not consider the Majorana phases. 
	We will only assume the charged lepton masses  as known parameters 
	and look for any possible hint into the spectra of neutrino masses. 
	Again, as in the quark sector, the charged lepton masses satisfy the same 
	hierarchical pattern, $m_e^2 \ll m_\mu^2 \ll m_\tau^2$, see Appendix \ref{app:masses}.
	From the three ratios, the largest one is $m_\mu/m_\tau
	\sim 10^{-2}$. So we safely neglect the other two ratios by taking the limit $m_e \rightarrow 0$. The 
	Pontecorvo--Maki--Nakagawa--Sakata (PMNS) matrix is then estimated as,
	 \begin{equation} \label{eq:pmnsMRP}
	 	\mathbf{U}_\text{PMNS} \simeq 
	 	\begin{pmatrix}
		1 & 0 & 0 \\
		0 & 1  & \Theta_{23}^\ell \\
		0 & -\Theta_{23}^\ell  & 1 \\
		\end{pmatrix}	
		\begin{pmatrix}
			c_{12}^\nu  c_{13}^\nu & -s_{12}^\nu c_{13}^\nu & -s_{13}^\nu e^{i\delta_\text{CP}^\nu} \\
			s_{12}^\nu c_{23}^\nu - c_{12}^\nu s_{23}^\nu s_{13}^\nu e^{-i\delta_\text{CP}^\nu} &
			c_{12}^\nu c_{23}^\nu + s_{12}^\nu s_{23}^\nu s_{13}^\nu e^{-i\delta_\text{CP}^\nu} & -s_{23}^\nu c_{13}^\nu \\
			s_{12}^\nu s_{23}^\nu + c_{12}^\nu c_{23}^\nu s_{13}^\nu e^{-i\delta_\text{CP}^\nu} &
			c_{12}^\nu s_{23}^\nu - s_{12}^\nu c_{23}^\nu s_{13}^\nu e^{-i\delta_\text{CP}^\nu} & c_{23}^\nu c_{13}^\nu
		\end{pmatrix} \; .
	 \end{equation}
	 The latter matrix product will only slightly modify the second and third rows of the unitary neutrino matrix. 
	 From which we can find and predict new mixing sum rules,
	 \begin{equation} \label{eq:mixsumrulLep}
	 \tan \Theta^\nu_{12} \simeq \tan \theta^\text{PMNS}_{12} \; , \quad
	 \sin \Theta^\nu_{13} \simeq \sin \theta^\text{PMNS}_{13} \; , \quad
	 \frac{-\Theta_{23}^\ell  + \tan \Theta^\nu_{23}}{1+\Theta_{23}^\ell \tan \Theta^\nu_{23}} \simeq \tan \theta_{23}^\text{PMNS} \; .
	 \end{equation}
	For more examples on mixing sum rules we refer the interested reader to~Refs.~\cite{Antusch:2009hq,Dorame:2012zv,Girardi:2014faa,
	Buccella:2017jkx,Gehrlein:2016fms,Gehrlein:2017ryu,Delgadillo:2018tza}.
	 
	In order to reproduce the observed values in lepton mixing, from Eqs.~\eqref{eq:LepMixingHierarchy} and \eqref{eq:pmnsMRP}, we conclude that neutrino masses should satisfy the following constraints:
	\begin{itemize}
		\item The lightest neutrino mass 
		cannot be zero, $m_{\nu 1} \neq 0$, as otherwise we should 
		simultaneously have very small solar and reactor mixing angles which 
		is not the case. In fact, through the following general relations,
		\begin{align}
		m_{\nu 1} = & \sqrt{\Delta m^2_{21}} \sqrt{\frac{x}{1-x}} \; , 
		\quad m_{\nu 2} = \sqrt{\Delta m^2_{21}} \sqrt{\frac{1}{1-x}} \; , \quad
		 m_{\nu 3} = 
		 \sqrt{\Delta m^2_{31} + \Delta m^2_{21} \frac{x}{1-x}} \;,  
		\end{align}
		where $x = m_{\nu 1}^2 / m_{\nu 2}^2$, we 
		can obtain a lower bound on the lightest neutrino mass,
		\begin{align}
			 m_{\nu 1} \geqslant  (8.6 \pm 0.1) \text{ meV}\; , 
		\end{align}
		where we have chosen $x=0.50$ as it is the largest value fulfilling both
		 $m_{\nu 1} < m_{\nu 2}$ and $m_{\nu 1}^2 \ll m_{\nu 2}^2$.\footnote{Here we are using the following
		 criteria. For $z \geqslant 0.5$ or $z < 0.5$ we approximate them as $z \sim {\cal O}(1)$ or $z \sim {\cal O}(10^{-1})$, respectively. In this way, if we want to have $m_1< m_2$ but at the same time $m_1^2 \ll m_2^2$, we can choose
		 the ratio, $x \equiv \tfrac{m_1^2}{m_2^2}$, to lie in the range $0.25 \leqslant x \leqslant0.50$ (or equivalently, $0.50 \leqslant \sqrt{x} \leqslant0.71$). } One may still soften the bound
		 if one chooses the smallest possible value for the ratio, $x=0.25$, obtaining,
		 \begin{align}
			 m_{\nu 1} \geqslant  (5.0 \pm 0.1) \text{ meV}\; . 
		\end{align}
				
		 \item Hierarchies among neutrinos cannot be as strong as with the charged fermions. 
		 Therefore, we may only speak of a hierarchy in the sense of $m_{\nu 3}^2 \gg m_{\nu 2}^2 \gg m_{\nu 1}^2$. In contrast to the charged fermion masses, which also satisfy $m_3 \gg m_2 \gg m_1$. 
	\end{itemize}	 	 
	 
	 Hence, from the observed values of the leptonic mixing matrix, $|\mathbf{U}_{\alpha k}| \gtrsim |\mathbf{V}_\text{us}|$ ($\alpha = e, \mu, \tau$, $k=1,2,3$), 
	 it is evident that neutrino masses should follow a rather different pattern from the charged fermion ones. Here again, if $\Theta_{13}$ is related to $\tfrac{m_{1}}{m_{3}}$, as it is the smallest
	 ratio, it should be the smallest mixing matrix element. 
	 Thus, in general, we shall theoretically always expect, through the MRP, the
	 $1-3$ element of the quark and lepton mixing matrices to be the smallest one.

	\paragraph{$\mu-\tau$ reflection symmetry.}
	Let us introduce a $\mu-\tau$ reflection symmetry in the neutrino sector~\cite{Harrison:2002et}, 
	$\tan \theta^\nu_{23} = 1$. Through the known values,
	$|\mathbf{U}_{\mu 3}| = 0.656$ and $|\mathbf{U}_{\tau 3}| = 0.739$, we can estimate the contribution coming from
	the charged lepton matrix,
	\begin{align}
		\Theta_{23}^\ell  \simeq 0.059 \;,
	\end{align}	 
	where we used the corresponding sum rule in Eq.~\eqref{eq:mixsumrulLep}.
	
	Curiously enough, the same value may be reached through the ratio $\tfrac{m_\mu}{m_\tau} = 0.059$. This
	meaning that the reflection symmetry can be easily cured by adding a rotation equal to the previous
	ratio, $\Theta_{23}^\ell  = \tfrac{m_\mu}{m_\tau}$. This could be seen as a 
	first hint on how individual mixing angles	could be related to mass ratios.

	\subsection{Effective Majorana mass}
We now apply the MRP to searches for the massive nature of neutrinos. 	
A clear signal of neutrinos as their own antiparticles may be reached through the study of processes where total lepton number is violated. In this sense,  neutrinoless double beta decay ($0\nu \beta\beta$), where total lepton number is violated by two units, offers a rare decay
to not only unveil the true massive nature of neutrinos but also to test predictions of left-right
symmetric models and other models including right-handed currents and heavy neutral leptons~\cite{Ge:2015yqa,Helo:2015ffa}. This rare process consists in an atom decaying into another one with the emission of two electrons,
\begin{equation}
(A,Z)\rightarrow (A,Z+2)+2e^-\;,
\end{equation} 
where ($A,Z$) are the mass and charge number, for the present status in these experiments see Ref.~\cite{Maneschg:2017mzu}. 

The study of this decay is made by the effective mass parameter $\langle m_{ee} \rangle$ which, in the
PDG parametrization, is expressed as,
\begin{align}
\begin{split}
\langle m_{ee} \rangle& =|\sum_j U^2_{ej}m_{\nu j}| \\
& =|m_{\nu 1} e^{2i\alpha}\cos^2\theta_{12}^\text{PMNS} \cos^2\theta_{13}^\text{PMNS} +m_{\nu 2} e^{2i\beta}\sin^2\theta_{12}^\text{PMNS}\cos^2\theta_{13}^\text{PMNS}+m_{\nu 3} e^{2i\delta}\sin^2\theta_{13}^\text{PMNS}|\;,
\end{split}
\end{align} 
where $\alpha$ and $\beta$ are the Majorana phases and $\delta$ is the Dirac phase.

Recently, the Cryogenic Underground Observatory for Rare Events (CUORE) which makes use of TeO${}_2$ crystals~\cite{Artusa:2014lgv} did not find any evidence for this decay in a limit on the effective Majorana neutrino mass $\langle m_{ee} \rangle < (0.11-0.52) \text{ eV}$~\cite{Alduino:2017ehq}. On the other hand, the GERmanium Detector Array (GERDA) which uses high purity germanium detectors enriched
with $^{76}$Ge~\cite{Agostini:2013mzu} have also excluded the range $\langle m_{ee}\rangle < (0.12-0.26)$ eV~\cite{Agostini:2018tnm}. The Enriched Xenon Observatory (EXO) experiment uses as a source and detector a pressurized time projection chamber filled with liquid Xenon~\cite{Albert:2014awa}. In a first stage EXO-200 has established the limit $\langle m_{ee}\rangle <(0.15-0.40)$ eV where still no evidence of the rare decay has been seen~\cite{Albert:2017owj}. The KAMioka Liquid Acintillator Anti-Neutrino Detector (KamLAND) is a multi-purpose detector that recently started the KamLAND-Zen experiment which in its second phase has reached the best measured limit so far, $ \langle m_{ee} \rangle <(0.06-0.16)$ eV~\cite{KamLAND-Zen:2016pfg}. 

Application of the MRP along with its three limits gives the possiblity
to study different benchmark scenarios for the effective mass parameter,
\begin{itemize}
	\item $\{m_1\}= \{ m_e, m_{\nu 1} \} \rightarrow 0$,
	\begin{equation}
		\langle m_{ee}^\text{MRP} \rangle  = 0 \; ,
	\end{equation} 
	compared to using any parametrization unconnected to the masses where one would get,
	\begin{equation}
		\langle m_{ee} \rangle  = |U^2_{e2}m_{\nu 2} + U^2_{e3}m_{\nu 3}| \; .
	\end{equation}
	From our mixing sum rules in Eq.~\eqref{eq:mixsumrulLep} we know that it
	 is safe to neglect the electron mass. Therefore, not having evidence for the
	 ocurrence for this decay would not necessarily mean, within the MRP,
	 that neutrinos are not Majorana. This would only mean the lightest neutrino mass
	 to be zero (or very small). 
	 An interesting observation that one may not reach from other parametrizations.
	Of course, from lepton mixing we have already concluded that $m_{\nu 1} \neq 0$, so this
	scenario is forbidden by the observed values in mixing.
	
	\item $\{m_1, m_2\}= \{ m_e, m_{\nu 1}, m_\mu, m_{\nu 2} \} \rightarrow 0$,
	\begin{equation}
		\langle m_{ee}^\text{MRP} \rangle  = 0 \; ,
	\end{equation} 
	in comparison to any other parametrization unconnected to the masses where one would get,
	\begin{equation}
		\langle m_{ee} \rangle  =  |U_{e3}|^2 m_{\nu 3} \; .
	\end{equation}
	This case is forbidden by the two mass squared differences, as they require at least two massive
	active neutrinos.
	
	\item $\{m_3\}= \{ m_\tau, m_{\nu 3} \} \rightarrow \infty$,
	\begin{equation}
		\langle m_{ee}^\text{MRP} \rangle  = |U^2_{e1}m_{\nu 1} + U^2_{e2}m_{\nu 2}| \; ,
	\end{equation} 
	compared to using any other parametrization unconnected to the masses where one would get,
	\begin{equation}
		\langle m_{ee} \rangle  \approx |U_{e3}|^2 m_{\nu 3} \; .
	\end{equation}
	We see that in any other parametrization we would conclude that the observed order of magnitude
	for the effective mass parameter would be approximately established by the heaviest mass
	whereas in the MRP would not play a role as it would have correctly decoupled. 
	
\end{itemize}

For last, we can compute the maximum value for the effective mass parameter as implied from the lower bounds
for the lightest neutrino mass and obtain,
\begin{align}
	\langle m_{ee}^{(x=0.25)} \rangle \leq 8.1 \text{ meV} \qquad \text{and} \qquad
	\langle m_{ee}^{(x=0.50)} \rangle \leq 11.4\  \text{ meV} \;,
\end{align}
values which still remain far from the experimental resolution.

\section{Conclusions}
\label{sec:Concl}

The flavour puzzle stands as the most intriguing set of still not understood aspects of the SM. All of them
originating from the fact that Nature has three fermion families. Here we have proposed and investigated
 the idea of connecting the mixing angles to the fermion mass ratios. That is, of building the concept
of a mixing parametrization whose four parameters are chosen to be the four independent mass
ratios in each fermion sector. By virtue of it, the observed values in the Cabibbo--Kobayashi--Maskawa (CKM) together with the ones appearing in the Pontecorvo--Maki--Nakagawa--Sakata (PMNS) matrices,
can be essentially understood. The strong hierarchical nature in the masses of the quark sector, translated into four very small ratios, gives as a consequence very small mixing angles. Thus, the closeness of the CKM matrix to the identity. On the other hand, even though the absolute scale of neutrino masses
is still unclear, through the same analysis we inferred that neutrino masses should have either a very mild hierarchy in the two ratios or at least in one of them, in order to produce such an anarchical structure in the PMNS matrix elements. 
In fact, in order for this analysis to work in the lepton sector, neutrino masses should have normal ordering and
the lightest neutrino mass should satisfy the lower bound
$m_{\nu 1} \geqslant (5.0\pm 0.1) \text{ meV}$ (for a more constrained scenario $m_{\nu 1} \geqslant (8.6\pm 0.1) \text{ meV}$), making the approach testable. 

Our approach has consisted in exploring the consequences of solely demanding 
that the mixing matrices
inherit the properties of the mass matrices under the limits of one, two, and/or three massless
fermion families, and the third family with an infinite mass. 
To this end, the naturalness criteria of 't Hooft~\cite{tHooft:1979rat} was taken into consideration as
the main argument to support this connection.  The properties to be fulfilled by the mixing matrices then
are: i) $\mathbf{V} = \mathbf{V} \left( \tfrac{m_u}{m_c},\tfrac{m_c}{m_t}, \tfrac{m_d}{m_s}, \tfrac{m_s}{m_b} \right)$, ii) $\mathbf{V} \left( 0,0,0,0 \right) = \mathbf{1}$,
iii) $ \mathbf{V} \left( 0,\tfrac{m_c}{m_t}, 0, \tfrac{m_s}{m_b} \right) = \mathbf{L}_{23}$,
iv) if $m_{t,b} \rightarrow \infty$ then $\mathbf{V} \left( \tfrac{m_u}{m_c},0,\tfrac{m_d}{m_s}, 0 \right) = \mathbf{L}_{12}$,
and a similar situation for the leptonic mixing matrix. 
For the cases ii) and iii) the minimal SM lagrangian acquires the corresponding global
symmetry $U(2)^6$ and $U(1)^6$ (or any continous or discrete subgroup), in consistency to the naturalness criteria of 't Hooft. The good agreement of the approach to the main aspects in quark and lepton mixing
can be taken as a strong evidence for the existence in Nature of a flavour symmetry.

\section*{Acknowledgments}
UJSS wants to acknowledge useful discussions with Lorenzo D\'iaz-Cruz and Wolfgang Gregor Hollik in the
initial stage of the work. 
The authors also want to acknowledge useful conversations 
with Florian Herren on the proper way to use \texttt{RunDec 3.0}.
The authors feel indebted to Stefano Morisi and Ivan Ni\v{s}and\v{z}i\'c for a careful reading of
the manuscript and their comments on it.
UJSS acknowledges support from a DAAD One-Year Research Grant. KMTN acknowledges
support from CONACYT-M\'exico. 
KMTN  feels very grateful to Konstantin Asteriadis and Florian Herren
for their warm hospitality during the realization and completion of this
work in their workspace and to the TTP members at KIT for their cordiality.

After completion of this work, the authors gratefully acknowledge receiving critical
insights and observations to the manuscript from Antonio Enrique C\'arcamo Hern\'andez, Wolfgang Gregor Hollik,
and Martin Spinrath. We thus thank them for carefully reading our manuscript and their comments on it.

\appendix
	
\section{Present status in fermion mixing}
\label{app:Mixing}
The most recent global fit from the PDG for the updated values of the Cabibbo--Kobayashi--Maskawa (CKM) mixing matrix
shows~\cite{Patrignani:2016xqp},
\begin{align}
|\mathbf{V}_\text{CKM}| =
   \begin{pmatrix}
	0.97434^{+0.00011}_{-0.00012}& 0.22506\pm 0.00050& 0.00357\pm 0.00015\\
0.22492\pm 0.00050&0.97351\pm 0.00013& 0.0411\pm 0.0013\\
0.00875^{+0.00032}_{-0.00033}& 0.0403\pm 0.0013& 0.99915\pm 0.00005\\
		\end{pmatrix} \; ,
\end{align}		
with the Jarlskog invariant equal to $ J_{\text{CKM}} =
(3.04^{+0.21}_{-0.20})\times 10^{-5}$. This set of numbers can be summarized by virtue of the standard parametrization,
\begin{align}
  \sin\theta_{12}^{\text{CKM}} = 0.22506 \pm 0.00050 \;,\qquad
  \sin\theta_{13}^{\text{CKM}} = 0.00357 \pm 0.00015\;, \\
  \sin\theta_{23}^{\text{CKM}} = 0.0411 \pm 0.0013\;, \qquad \text{ and } \qquad
  \delta_{\text{CP}}^{\text{CKM}} = (71.6^{+1.3}_{-1.0})^{\circ}.
\end{align}
Whereas the fit for the improved Wolfenstein parameters gives\cite{Patrignani:2016xqp},
\begin{align}
        \lambda = 0.22506 \pm 0.00050 \; , \quad\quad A= 0.811\pm 0.026 \;, \\
        \bar{\rho} = 0.124^{+0.019}_{-0.018} \;, \quad\quad \bar{\eta} = 0.356 \pm 0.011 \; .
\end{align}	

On the other hand, the current global fit values for the Pontecorvo--Maki--Nakagawa--Sakata (PMNS) mixing matrix at $3\sigma$ are~\cite{Esteban:2016qun}:
\begin{align}
|\mathbf{U}_\text{PMNS}| =
   \begin{pmatrix}
0.799\rightarrow 0.844& 0.516\rightarrow 0.582 & 0.141\rightarrow 0.156\\
0.242\rightarrow 0.494& 0.467\rightarrow 0.678 & 0.639\rightarrow 0.774\\
0.284\rightarrow 0.521 &0.490\rightarrow 0.695 & 0.615\rightarrow 0.754
\end{pmatrix} \;,
\end{align}	
with the Jarlskog invariant at $1\sigma$ as $J_\text{PMNS}^\text{max}=-(0.0329 \pm 0.0007)$. When expressed in the standard parametrization~\cite{Esteban:2016qun},
\begin{align}
  \sin^2\theta_{12}^{\text{PMNS}} = 0.307^{+0.013}_{-0.012} \; ,\qquad
  \sin^2\theta_{13}^{\text{PMNS}} = 0.02206 \pm 0.00075 \; , \\
  \sin^2\theta_{23}^{\text{PMNS}} = 0.538^{+0.033}_{-0.069} \; , \qquad \text{ and } \qquad
  \delta_{\text{CP}}^{\text{PMNS}} = (234^{+43}_{-31})^{\circ} \; .
\end{align}

\section{Present status in fermion masses}
\label{app:masses}

\begin{center}
\begin{tabular}{ p{5cm}||p{4.5cm} }
 \toprule[0.1em]
  \multicolumn{2}{c}{\textbf{QUARK MASSES}} \\
 \midrule[0.1em]
 \begin{center} Experimental masses \hspace{4cm} Input $(\text{GeV})$
  \end{center} & \begin{center} Masses at $M_z$ scale \hspace{4cm}      Output ($\text{GeV}$) \end{center}\\
 \hline
 \vspace{0.01cm}
 $m_u(2 \text{ GeV})=0.0022_{-0.0004}^{+0.0006}$  &  \vspace{0.1cm}$m_u(M_z)=0.0013_{-0.0002}^{+0.0003}$\\
 \vspace{0.01cm}
 $m_d(2 \text{ GeV})=0.0047_{-0.0004}^{+0.0005}$ &  \vspace{0.01cm}$m_d(M_z)=0.0027_{-0.0002}^{+0.0003}$\\
  \vspace{0.01cm}
 $m_s(2 \text{ GeV})=0.096_{-0.004}^{+0.008}$ &  \vspace{0.01cm}$m_s(M_z)=0.055_{-0.002}^{+0.004}$\\
  \vspace{0.01cm}
 $m_c(m_c)=1.27\pm0.03$ &  \vspace{0.01cm}$m_c(M_z)=0.626\pm0.02$\\
 \vspace{0.01cm}
 $m_b(m_b)=4.18_{-0.03}^{+0.04}$ &  \vspace{0.01cm}$m_b(M_z)=2.86_{-0.02}^{+0.02}$\\
  \vspace{0.01cm}
 $m_t(\text{OS})=173.21 \pm 0.87$ &  \vspace{0.01cm}$m_t(M_z)=172.29\pm 0.06$  \vspace{0.1cm}\\ 
 \bottomrule[0.1em]
\end{tabular}
\captionof{table}{Here we present in the left column the most recent measured masses 
as taken from \cite{Patrignani:2016xqp}. By virtue of the \texttt{RunDec} package they are run
to the $Z$ boson mass scale \cite{Herren:2017osy}. \texttt{RunDec} takes into account the five-loop corrections of the QCD beta function and four-loop effects when decoupling the heavy quarks below their energy scale.} \label{tab:mass1}
\end{center}
 
 \begin{center}
\begin{tabular}{ p{5cm}||p{5cm}  }
 \toprule[0.1em]
  \multicolumn{2}{c}{\textbf{LEPTON MASSES}} \\
 \midrule[0.1em]
 \begin{center} Charged lepton \hspace{4cm} ($\text{MeV}$)
 \end{center} & \begin{center} Neutrino mass differences \hspace{4cm}       ($\text{eV}^2$) \end{center}\\
 \hline
 \vspace{0.1cm}
 $m_e(M_z)= 0.4861410527$  &  \vspace{0.1cm}$\frac{\Delta m_{21}^2}{10^{-5}}=7.40^{+0.21}_{-0.20}$\\
 \vspace{0.01cm}
 $m_\mu(M_z)=102.627051$ &  \vspace{0.01cm}IO: $\frac{\Delta m_{32}^2}{10^{-3}}=-2.465^{+0.032}_{-0.031}$\\
  \vspace{0.01cm}
 $m_\tau(M_z)=1744.614156$ &  \vspace{0.01cm}NO: $\frac{\Delta m_{31}^2}{10^{-3}}=+2.494^{+0.033}_{-0.031}$  \vspace{0.1cm}\\
\bottomrule[0.1em]
\end{tabular}
\captionof{table}{This table presents the charged lepton masses as taken from Ref.~\cite{Antusch:2013jca} and the updated neutrino mass differences~\cite{Esteban:2016qun}. We have denoted by NO and IO the Normal and Inverted Ordering scenarios, respectively. We have omitted the experimental error from the charged leptons due to their size which makes no difference in the error propagation.}
\label{tab:mass2}
\end{center}	

\section{Mixing in the rank one case}
\label{app:Basis}

Without any loss of generality, let us consider the real case for simplicity. 
The two rank one mass matrices of a given fermionic sector are denoted as,
\begin{align}
	\widetilde{\bf M}^f \simeq & \begin{pmatrix}
		L^{f}_{31} R^f_{31} & L^{f}_{31} R^f_{32} & L^{f}_{31} R^f_{33} \\
		L^{f}_{32} R^f_{31} & L^{f}_{32} R^f_{32} & L^{f}_{32} R^f_{33} \\
		L^{f}_{33} R^f_{31} & L^{f}_{33} R^f_{32} & L^{f}_{33} R^f_{33}
	\end{pmatrix} \; ,
\end{align}
where $f=u,d\; (\nu,e)$ and we have normalized by the largest singular value, $\widetilde{\bf M}^f = {\bf M}^f/m_{f,3}$. 

We are interested in the left Hermitian product, $\widetilde{\bf M}^f \widetilde{\bf M}^f{}^{\dagger}$,
\begin{align}
	\widetilde{\bf M}^f \widetilde{\bf M}^f{}^{\dagger} = & \begin{pmatrix}
		L^{f}_{31} L^f_{31} & L^{f}_{31} L^f_{32} & L^{f}_{31} L^f_{33} \\
		L^{f}_{32} L^f_{31} & L^{f}_{32} L^f_{32} & L^{f}_{32} L^f_{33} \\
		L^{f}_{33} L^f_{31} & L^{f}_{33} L^f_{32} & L^{f}_{33} L^f_{33}
	\end{pmatrix} \; .
\end{align}
Its diagonalization occurs via the orthogonal transformation,
\begin{align}
	{\bf R}^f = & \begin{pmatrix}
		\frac{-L^{f}_{33}}{\sqrt{L^{f}_{31}{}^2+L^{f}_{33}{}^2}} &
		\frac{-L^{f}_{31} L^{f}_{32}}{\sqrt{L^{f}_{31}{}^2+L^{f}_{33}{}^2}} &
		L^{f}_{31} \\
		0 &
		\sqrt{L^{f}_{31}{}^2 + L^{f}_{33}{}^2} &
		L^{f}_{32} \\
		\frac{-L^{f}_{31}}{\sqrt{L^{f}_{31}{}^2+L^{f}_{32}{}^2}} &
		\frac{-L^{f}_{32} L^{f}_{33}}{\sqrt{L^{f}_{31}{}^2+L^{f}_{33}{}^2}} &
		L^{f}_{33} \\
	\end{pmatrix} \; ,
\end{align}
where we are assuming the singular vector is already normalized.

In the mass basis, fermionic mixing is computed from,
\begin{align}
	{\bf W} = & {\bf R}^a {\bf R}^b{}^\dagger \; .
\end{align}
Now, in general, this will give us a $3 \times 3$ mixing matrix with non-zero off-diagonal elements.
This conclusion remains even if we explicitly consider the possibility of rotating the $1-2$ sector with a possible
$U(2)_L^{F=Q,E}$ symmetry. Hence, taking the massless limit for the two first families will still imply fermionic mixing
and emergent global symmetry factors will not be possible.

\bibliographystyle{utcaps}		
\bibliography{bib}

\end{document}